\documentclass[showpacs,reprint,amsmath,amssymb,aps,prb,floatfix,twocolumn,superscriptaddress]{revtex4}

\usepackage{graphicx}
\usepackage{dcolumn}
\usepackage{bm}
\usepackage{verbatim}
\usepackage{color}

\newcommand{\dd}{\mathrm{\mathrm{d}}}

\begin{document}

\title{Symmetry-breaking, spin- and orbital-freezing in the two-band Hubbard model}
\title{Long-range orders, spin- and orbital-freezing in the two-band Hubbard model}
\author{Karim Steiner}
\affiliation{Department of Physics, University of Fribourg, 1700 Fribourg, Switzerland}
\author{Shintaro Hoshino}
\affiliation{Department of Basic Science, The University of Tokyo, Meguro, Tokyo 153-8902, Japan}
\author{Yusuke Nomura}
\affiliation{Centre de Physique Th\'eorique, \'Ecole Polytechnique, CNRS, Universit\'e Paris-Saclay, F-91128 Palaiseau, France}
\author{Philipp Werner}
\affiliation{Department of Physics, University of Fribourg, 1700 Fribourg, Switzerland}

\date{\today}

\begin{abstract}
We solve the orbitally degenerate two-band Hubbard model within dynamical mean field theory and map out the instabilities to various symmetry-broken phases based on an analysis of the corresponding lattice susceptibilities. Phase diagrams as a function of the Hund coupling parameter $J$ are obtained both for the model with rotationally invariant interaction and for the model with Ising-type anisotropy. For negative $J$, an intra-orbital spin-singlet superconducting phase appears at low temperatures, while the normal state properties are characterized by an orbital freezing phenomenon. This is the negative-$J$ analogue of the recently discovered fluctuating-moment induced $s$-wave spin-triplet superconductivity in the spin-freezing regime of multi-orbital models with $J>0$.  
\end{abstract}


\maketitle

\section{Introduction} 

The Hund coupling $J$ leads to interesting correlation effects in multi-orbital Hubbard models.\cite{Georges2013} For a fixed total occupation of a given site, the interaction $J$ differentiates between the energies of different orbital occupation and spin states. In a lattice environment, this can lead to considerable shifts in the metal-Mott insulator phase boundaries,\cite{Inaba2005,Werner2009} to local moment formation,\cite{Werner2008} nonlocal correlation effects,\cite{PhysRevB.79.245128,PhysRevB.91.235107} and various types of symmetry-breaking.\cite{Karsten_ferro,PhysRevB.58.R567,PhysRevLett.99.216402,Chan2009,Sakai2004,Han2004,Hoshino2015} Already in the two-band case, nontrivial crossovers and phase transitions can be observed. In the presence of a crystal field splitting between the bands, the strength of the Hund coupling controls the competition between high-spin and low-spin solutions, and induces a correlated metallic state between the corresponding insulating phases.\cite{Werner2007_crystal} This state has been found to be unstable with respect to spin-orbital ordering,\cite{Hoshino2016} or, in an alternative language, to excitonic condensation.\cite{Kunes2014} But even the orbitally degenerate system displays a range of interesting phenomena. For example, in the disordered metallic phase away from half-filling, a spin-freezing crossover\cite{Werner2008} occurs at  specific values of the filling and interaction.\cite{Medici2011,Hafermann2012} At low temperature, fluctuating local moments at the border of the spin-frozen regime lead to an orbital-singlet spin-triplet superconducting instability.\cite{Hoshino2016} This superconducting phase borders an antiferromagnetic phase near half-filling, and in the doped system 
a ferromagnetic phase at large $U$. Intra-orbital spin-singlet pairing occurs in models with a negative $J$.\cite{Koga2015,Capone_2002} 

Motivated by this rich physics, we undertake here a systematic study of the various instabilities to long-range ordered phases by computing the lattice susceptibilities for homogeneous and staggered order parameters. We focus on the orbitally degenerate model, but consider both the rotationally invariant interaction, and the Ising anisotropic interaction of density-density type.
We map out the finite-temperature phase diagrams as a function of $J$ for fixed $U$, to emphasize the role of the Hund coupling as the key player responsible for most of the instabilities and crossovers. We do not restrict our study to the usual range $0<J<U/3$, but also consider negative Hund couplings (relevant for example in connection with the physics of alkali-doped fullerides),\cite{Capone_2002,RevModPhys.81.943,nomura_C60_paper,nomura_C60_review,PhysRevB.85.155452,PhysRevB.92.245108} and $J>U/3$, which has been considered in discussions of the metal-insulator transition and orbital ordering phenomena in nickelates.\cite{PhysRevLett.98.176406,PhysRevB.91.075128,Rupen2015}

The rest of this paper is organized as follows: Section~\ref{sec:method} details the model and the method used to compute the lattice susceptibilities, Section~\ref{sec:half} presents the phasediagrams of the half-filled model in the density-density approximation and for the rotationally invariant system. Section~\ref{sec:38} discusses the system at 3/8 filling and the orbital freezing phenomenon for negative $J$. Section~\ref{sec:outlook} is a conclusion and outlook.

\section{Model and method} 
\label{sec:method}

We consider the two-orbital Hubbard model on an infinite-dimensional Bethe lattice. The Hamiltonian of the lattice model is given by
\begin{align}
H=&-t\sum_{\langle i,j\rangle,\alpha,\sigma} d^\dagger_{i,\alpha\sigma}d_{j,\alpha\sigma}-\mu\sum_{i,\alpha,\sigma} \dd \tau n_{i,\alpha\sigma}\nonumber\\
&\hspace{28mm}+ \sum_i (H_{\text{int},i}^\text{dens}+H_{\text{int},i}^\text{sf-ph}),
\end{align}
where $i$ denotes the site, $\alpha$ the orbital, $\sigma$ the spin, $t$ 
the hopping parameter and $\mu$ the chemical potential. $n_{i,\alpha\sigma}=d^\dagger_{i,\alpha\sigma}d_{i,\alpha\sigma}$ is the spin- and orbital-dependent density on site $i$,  $H_{\text{int},i}^\text{dens}$ represents the density-density part of the Slater-Kanamori interaction, and $H_{\text{int},i}^\text{sf-ph}$ the spin-flip and pair-hopping terms. In the infinite-dimensional limit, and with a rescaling $t\rightarrow t^*/\sqrt{d}$, this model can be solved exactly within dynamical mean field theory (DMFT).\cite{Metzner1989,Georges1992} The DMFT formalism maps the lattice problem onto a self-consistent solution of an impurity model with action
\begin{align}
S_\text{imp} =& \iint_0^\beta d\tau d\tau' \sum_{\alpha,\sigma} d^\dagger_{\alpha\sigma}(\tau)\Delta_{\alpha\sigma}(\tau-\tau')d_{\alpha\sigma}(\tau')\nonumber\\
&\hspace{0mm}-\mu\int_0^\beta d\tau \sum_{\alpha,\sigma} n_{\alpha\sigma}+\int_0^\beta d\tau (H_\text{int}^\text{dens}+H_\text{int}^\text{sf-ph}),\\
H_\text{int}^\text{dens} =& \sum_\alpha U n_{\alpha\uparrow} n_{\alpha\downarrow} +\sum_{\sigma} U' n_{1\sigma} n_{2\bar\sigma} \nonumber\\
&\hspace{0mm}+  \sum_{\sigma} (U'-J) n_{1\sigma} n_{2\sigma}, \\
H_\text{int}^\text{sf-ph} =& - J (d_{1\downarrow}^{\dagger} d_{2\uparrow}^{\dagger} d_{2\downarrow} d_{1\uparrow} + d_{2\uparrow}^{\dagger} d_{2\downarrow}^{\dagger} d_{1\uparrow} d_{1\downarrow})+{\rm h.c.}
\end{align}
We use the relation $U'=U-2J$ between the interorbital repulsion $U'$ and the intraorbital repulsion $U$, which is valid for rotationally invariant systems.
The orbitally diagonal hybridization function $\Delta_{\alpha\sigma}$ is fixed by the DMFT self-consistency condition, which for the infinite-dimensional Bethe lattice with bandwidth $4t^*$ simplifies to\cite{Georges_1996} 
\begin{equation}
\Delta_{\alpha\sigma}(\tau)=(t^*)^2 G_{\alpha\sigma}(\tau),
\end{equation} 
with $G_{\alpha\sigma}(\tau)=
-
\text{Tr}_d[{\mathcal T}e^{-S_\text{imp}}d_{\alpha\sigma}(\tau)d^\dagger_{\alpha\sigma}(0)]/\text{Tr}[{\mathcal T}e^{-S_\text{imp}}]$ the impurity Green's function. In this study, we consider such a Bethe lattice, and use $t^*$ as the unit of energy.

We solve the impurity model with the recently developed double-expansion impurity solver,\cite{Steiner2015} which combines a hybridization expansion\cite{Werner_2006} with a weak-coupling expansion in $H^\text{sf-ph}$, and allows an efficient simulation of the two-orbital model with and without the spin-flip and pair-hopping terms. To find the potential stability regions of long-range ordered phases, we compute the corresponding lattice susceptibilities $\chi$ and search for divergences in these susceptibilities. Within DMFT, the lattice susceptibilities are obtained by first extracting the local vertex from the four-point correlation functions of the impurity model, and a subsequent solution of the lattice Bethe-Salpeter equations. In the rotationally invariant case, symmetry relations can be used to express all required four-point correlation functions in terms of easily measurable ones. Detailed explanations can be found in Refs.~\onlinecite{Hoshino2015,Hoshino2016}.

\section{Results for half-filling}
\label{sec:half}

\subsection{Phase diagram}
 
\begin{figure*}[t]
	\includegraphics[width=8.7cm]{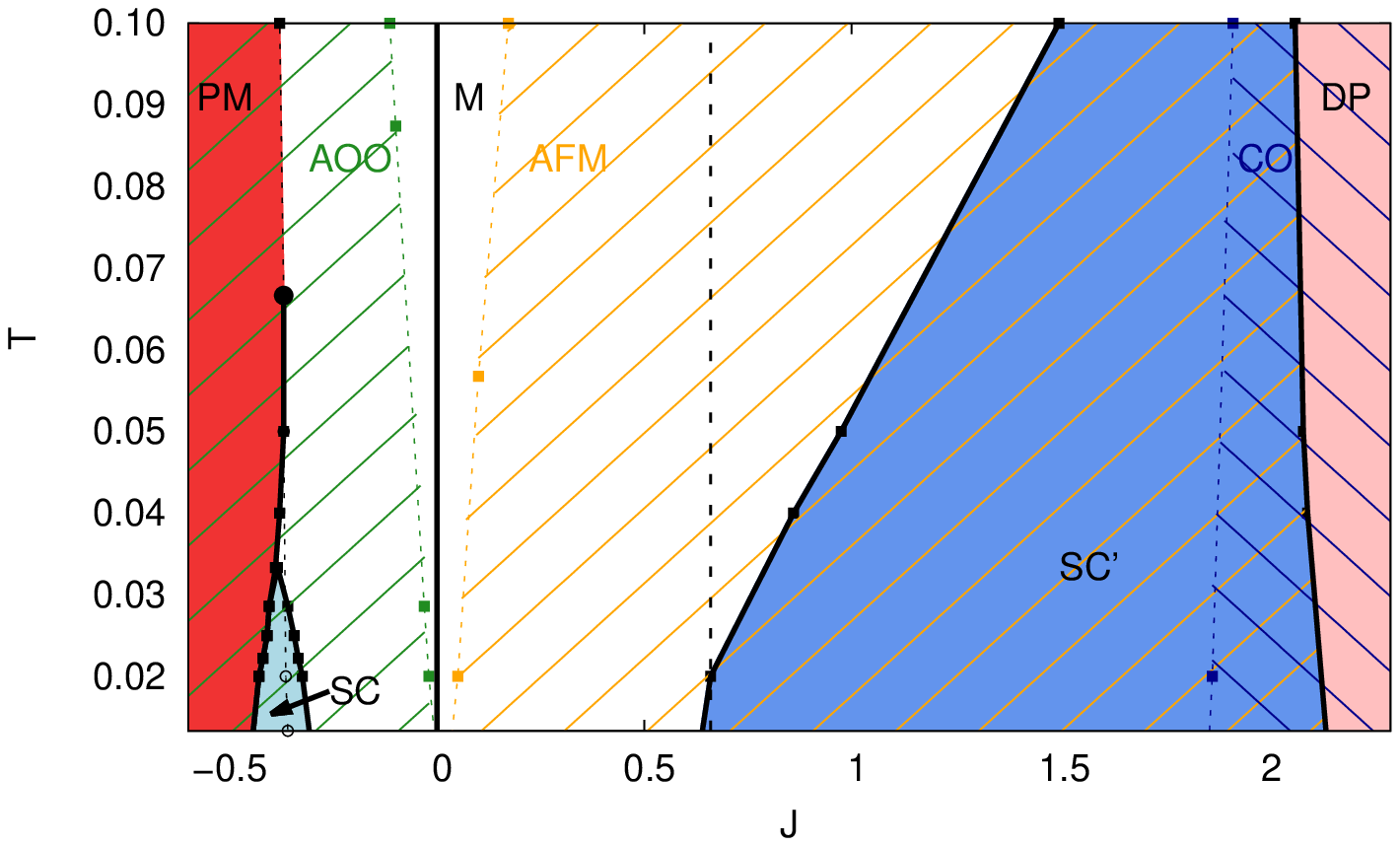}\hfill
	\includegraphics[width=8.7cm]{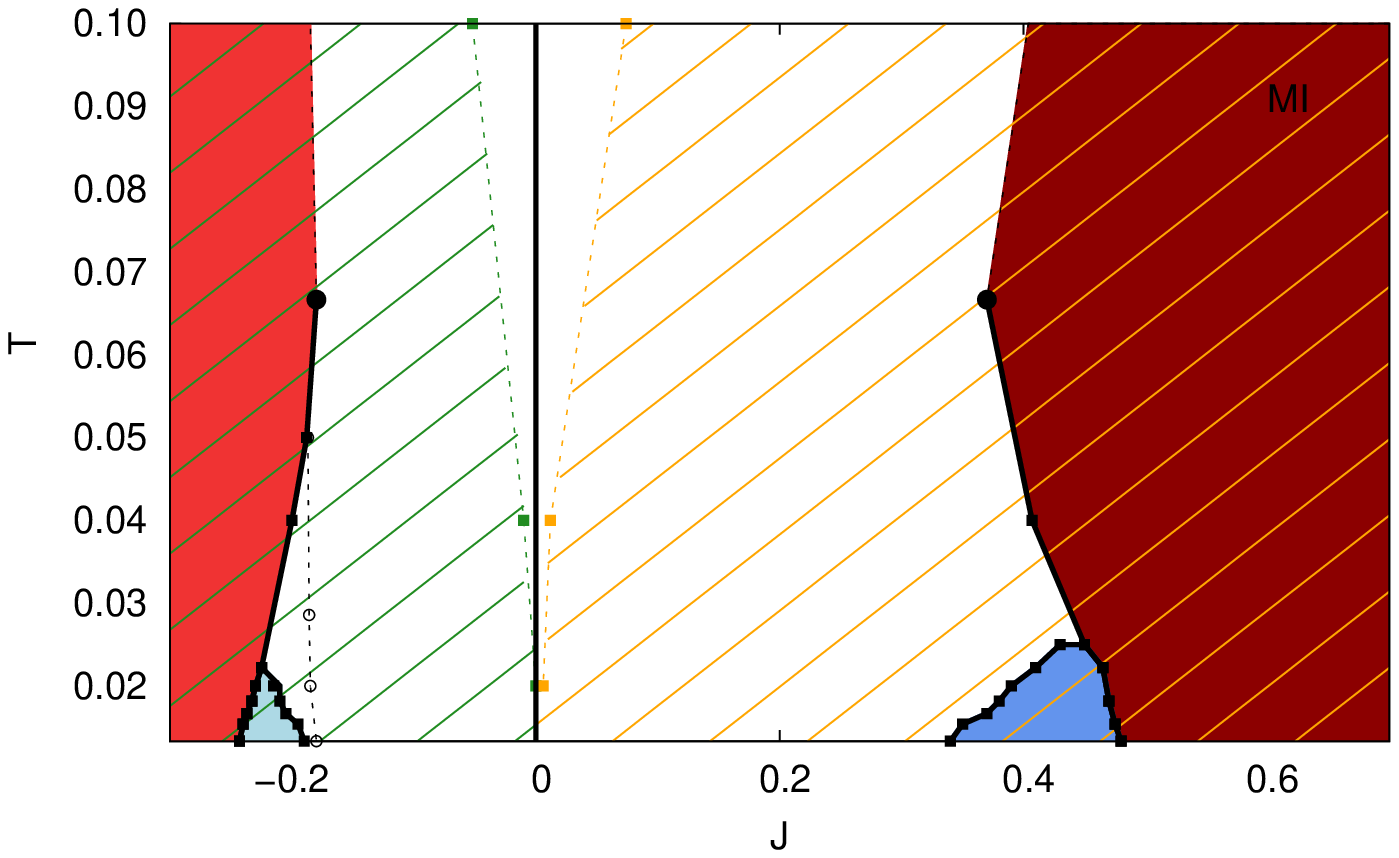}
	\includegraphics[width=8.7cm]{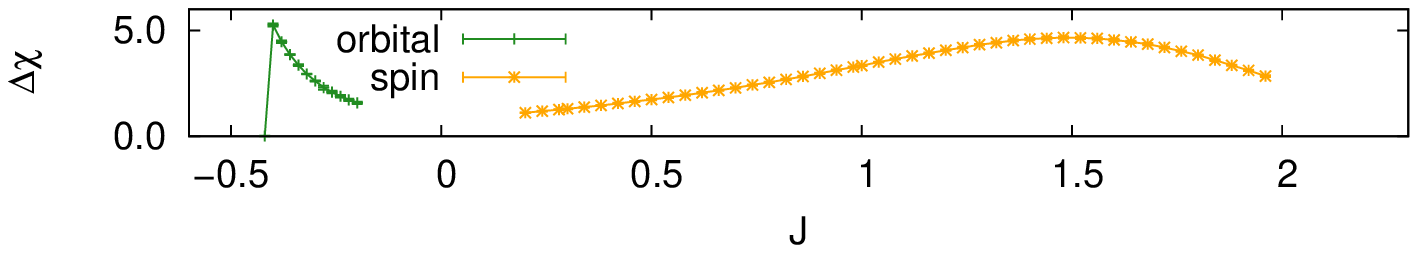}\hfill
	\includegraphics[width=8.7cm]{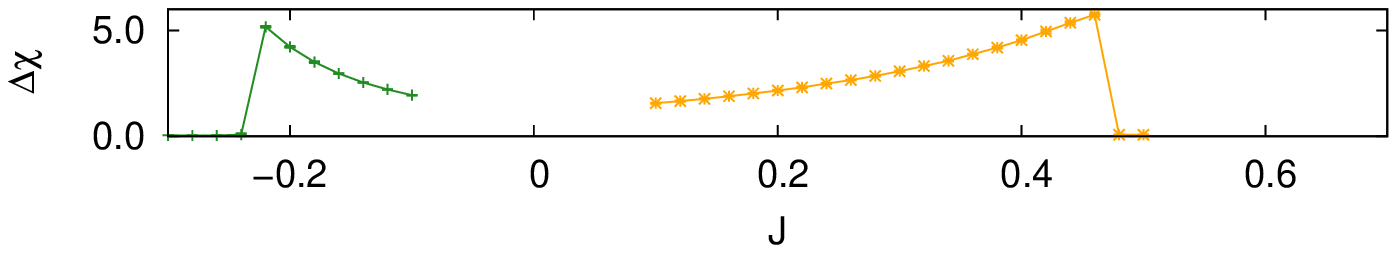}\
	\includegraphics[width=8.7cm]{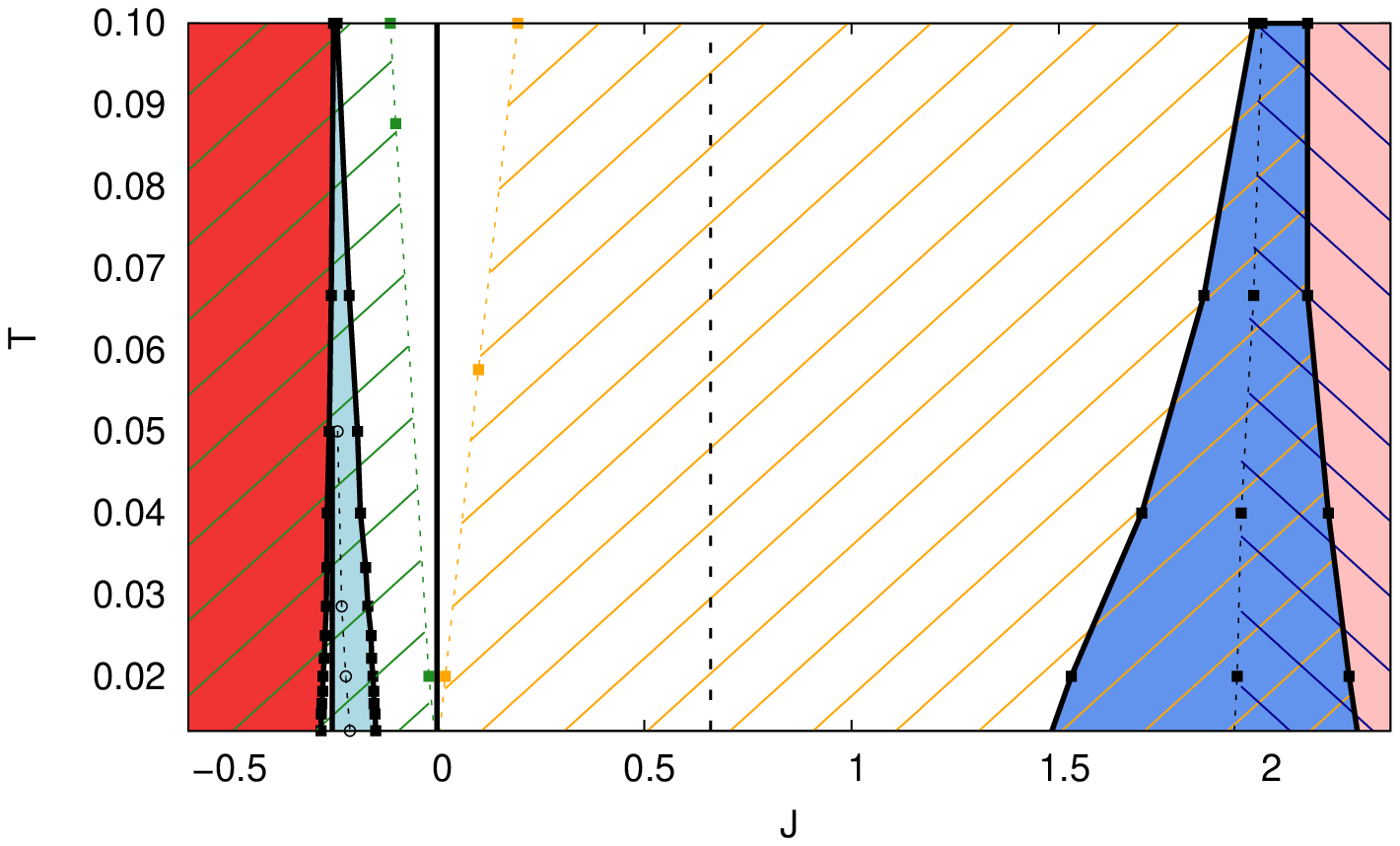}\hfill
	\includegraphics[width=8.7cm]{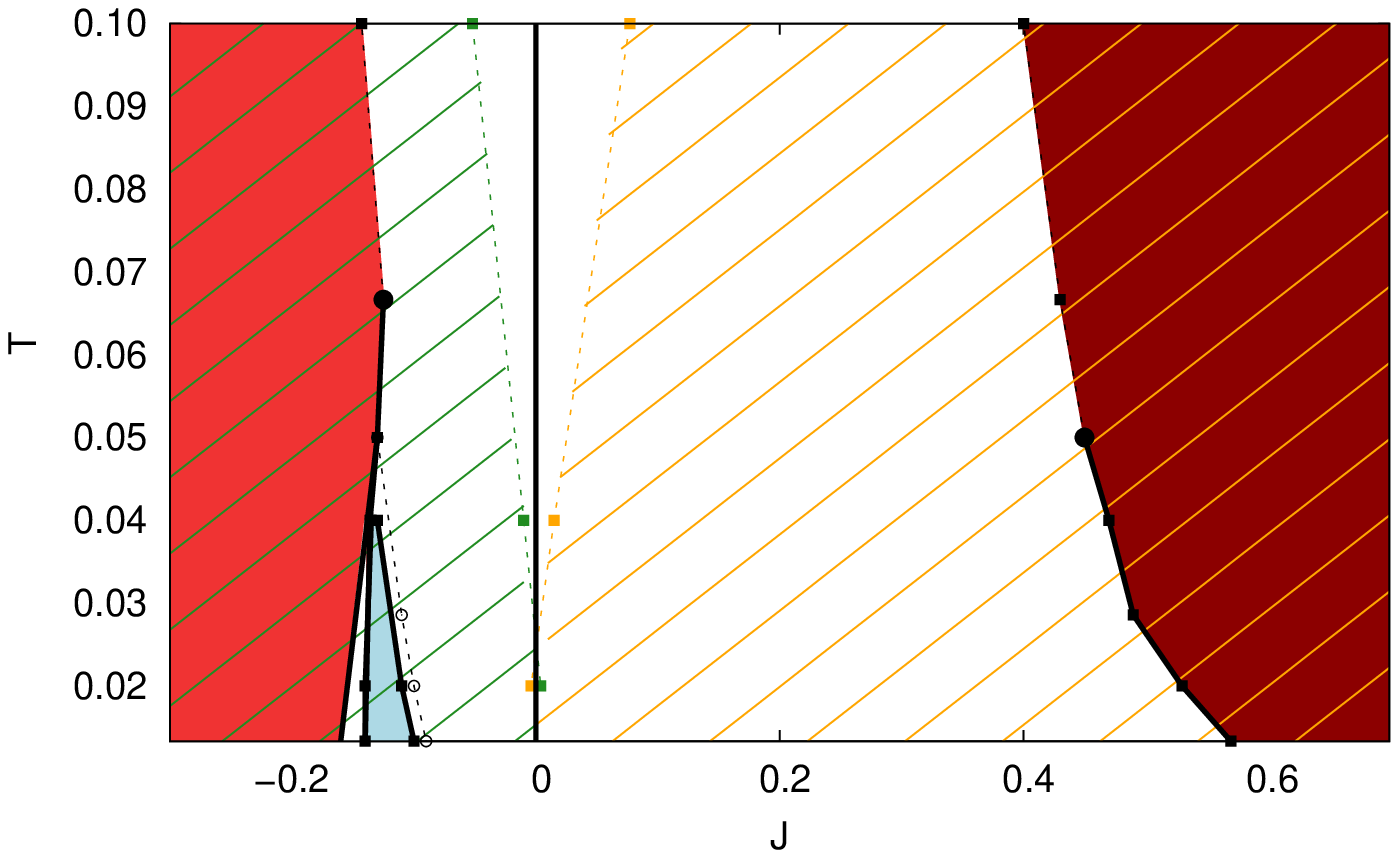}
	\includegraphics[width=8.7cm]{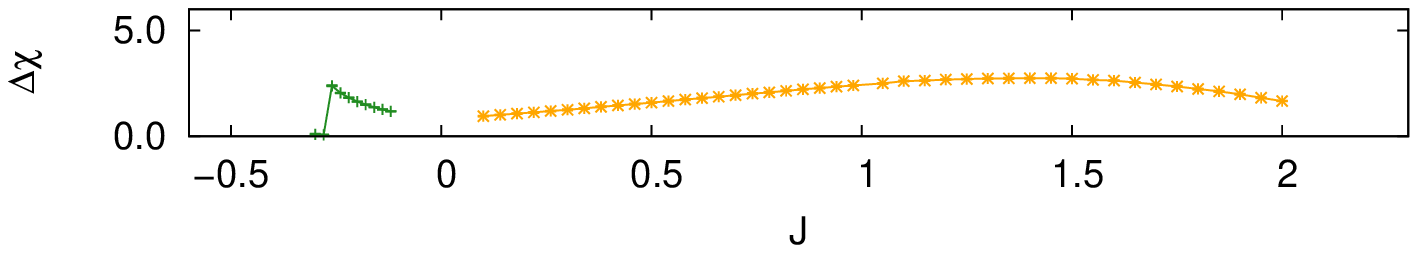}\hfill
	\includegraphics[width=8.7cm]{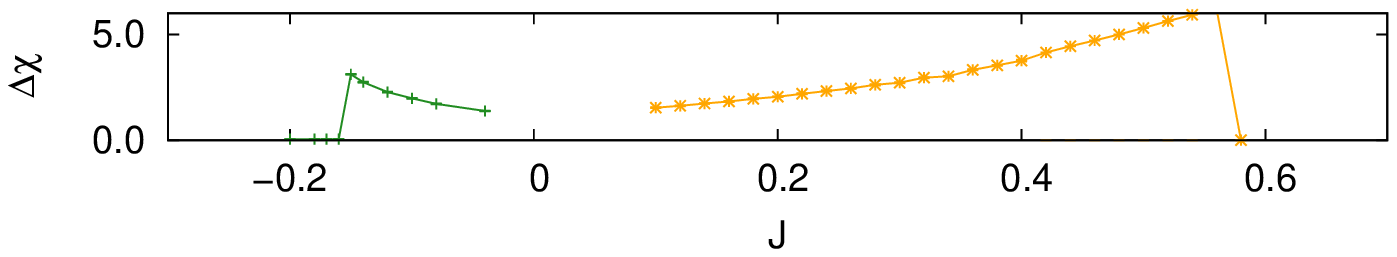}\
	\caption{\label{fig_half}
	(Color online) Phase diagrams of the half-filled system in the space of $J$ and $T$ for fixed $U=2$ (left panels) and $U=4$ (right panels). Both the results for Ising anisotropy (top panels) and the rotationally invariant interaction (bottom panels) are shown. The following phases are found: paired Mott insulator (PM), intra-orbital spin-singlet pairing (SC), metal phase (M), interorbital spin-triplet pairing (SC'), conventional Mott insulator (MI), and double paired state (DP).
	Antiferromagnetic order (AFM), antiferro orbital order (AOO), and charge order (CO) appear in the hashed regions of the phase diagram. 
	The solid metal-insulator boundaries have been obtained by starting from a metallic initial solution and thus correspond to $J_{c2}$. On the $J<0$ side, we indicate $J_{c1}$ (stability region of the insulator) by a dashed line with empty circles. 
	Dashed lines emanating from the end points of the metal-insulator transition lines indicate a metal-insulator crossover. 
The thin panels below the phase diagrams show the $J$-dependence of the local spin and orbital fluctuations for $\beta=75$ (see text).
	}
\end{figure*}

Figure~\ref{fig_half} shows phase diagrams of the half-filled system in the space of Hund coupling $J$ and temperature $T$ for fixed $U=2$ (left panels) and $U=4$ (right panels). The top panels are for the model with density-density interaction, and the bottom panels for the rotationally invariant interaction. For small $J$, the calculations, which are performed in a state without symmetry breaking, produce a metallic solution, while for large negative or positive $J$, we find first order transitions to a paired Mott insulator (PM), a conventional Mott insulator (MI), and a double-paired (low-spin) state (DP), respectively.
The different nature of the insulators becomes evident, for example, by looking at the histogram of atomic states, which is dominated by the orbital occupations $(n_1,n_2)=(2,0)$ and $(0,2)$  in the case of the paired Mott insulator (88\% 
for $U=4$, $J=-0.20$, $\beta=50$, rotationally invariant interaction), while the half-filled orbital states $(1,1)$ dominate the histogram in the conventional Mott insulator (88\% 
for $U=4$, $J=0.60$, $\beta=50$, rotationally invariant interaction). The double paired state DP, which appears near the region where the attractive inter-orbital opposite-spin interaction dominates the repulsive intra-orbital interaction 
($U+U'<0$, i.e. $J> U$), has dominant states with filling $(2,2)$ and $(0,0)$. 

A first order transition between metal and insulator is only found at sufficiently low temperature. Above a critical end-point, marked by a black dot in Fig.~\ref{fig_half}, the transition turns into a crossover, indicated by dashed lines. We have located the end-point by studying the $J$-dependence of $G(\beta/2)$ and looking for the disappearance of a 
jump. The crossover line corresponds to the inflection point of the $G(\beta/2)$-vs-$J$ curve.

At low temperature, in the vicinity of the paired Mott state, a diverging susceptibility for intra-orbital spin-singlet pairing indicates the existence of an $s$-wave spin-singlet superconducting phase (SC). 
In the vicinity of the $J>0$ insulators, except for the rotationally invariant case with $U=4$,
we find a different kind of $s$-wave superconducting instability, namely the appearance of an orbital-singlet spin-triplet phase (SC') analogous to the fluctuating-moment induced superconducting state discussed in the three-orbital context in Ref.~\onlinecite{Hoshino2015}.

Apart from two distinct $s$-wave superconducting phases, we also find instabilities to antiferromagnetic order (AFM, for $J>0$) and antiferro orbital order (AOO, for $J<0$) in the low-temperature region considered.  These orders are related to the activation of spin and the suppression of orbial degrees of freedom for $J>0$, and vice versa for $J<0$, which is consistent with the existence of spin-singlet and spin-triplet superconductivity for positive and negative $J$. The corresponding symmetry breaking would open a gap in the spectrum and  we would see a crossover from a Slater-mechanism-induced insulator to an insulator with well-developed local spin/orbial moments. At $J=0$ and for $T\ge 0.01$, 
there is no magnetic or orbital ordering in the $U=2$ case,  
so the ordered insulating phases are separated at small $|J|$ by a metallic region. 
In the phasediagrams for $U=4$, we find a crossing of the AOO and AFM instability lines at low temperatures near $J=0$, so there should be a first-order transition between the two insulating phases. 
The crossing of the AOO and AFM lines is expected to occur exactly at $J=0$, where the system has an SU(4) symmetry and AOO and AFM are degenerate.

In the $U=2$ case, there is furthermore an instability to charge order (CO) close to $J=U$. The CO region extends into the DP phase, while the AFM instability is confined to the metallic region. This is because the local spin moments are quenched in the DP phase.  
On the other hand, the MI in the phase diagram for $U=4$ consists of localized spin $S=1$ moments, which are susceptible to antiferromagnetic order. Hence the AFM region also covers the MI.

\begin{figure}
         \includegraphics[width=8.7cm]{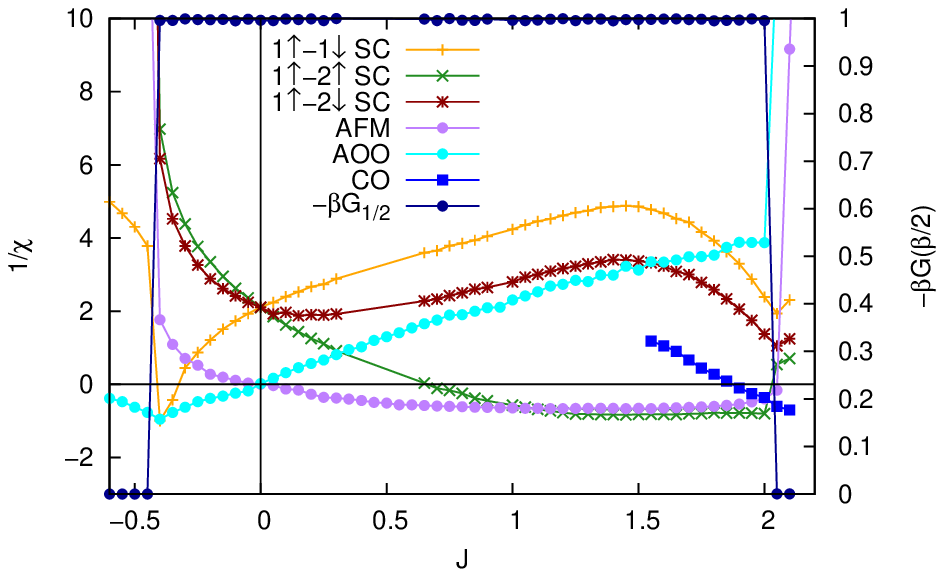}
	\includegraphics[width=8.7cm]{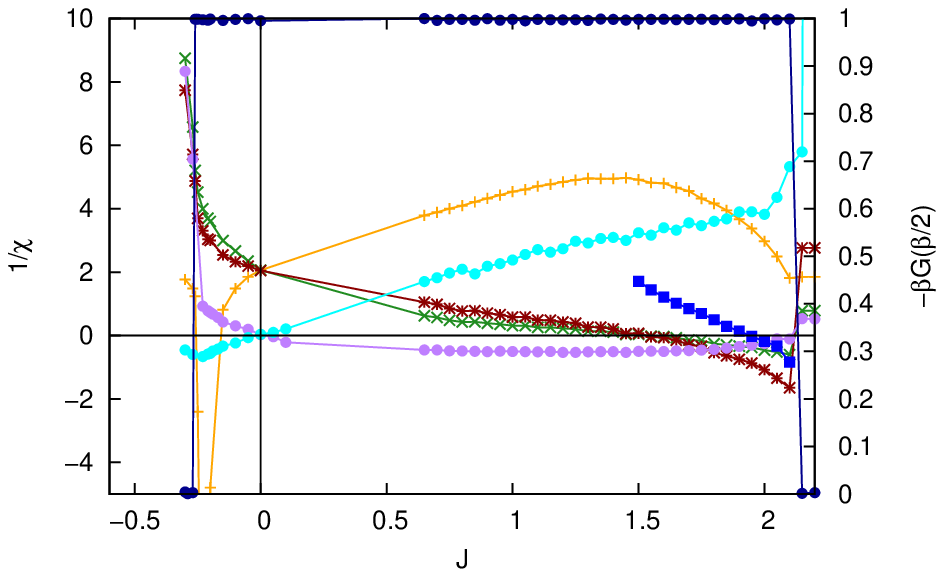}
	\caption{\label{fig_susc}
	Inverse susceptibilities for different pairings, AFM, AOO and CO as a function of $J$ for $U=2$ and $\beta=50$. The top panel shows the result for the model with Ising anisotropy, and the bottom panel for the model with rotationally invariant Hund coupling. 
	Negative values of inverse susceptibilities indicate a possible symmetry breaking.
	Also plotted is the estimate $-\beta G(\beta/2)$ for the density of states at the Fermi energy. 
}
\end{figure}

To illustrate the order of the transitions and the different nature of the two superconducting phases SC and SC', we plot in Fig.~\ref{fig_susc} the inverse pairing susceptibilities for different orbitals and spins. The susceptibilities are defined as
\begin{equation}
\chi_{\alpha\sigma, \beta\sigma'}=\frac{1}{N}\int_0^\beta \langle O_{\alpha\sigma, \beta\sigma'}(\tau) O^\dagger_{\alpha\sigma, \beta\sigma'}(0)\rangle d\tau,
\end{equation}
where $N$ is the number of sites, and $O_{\alpha\sigma, \beta\sigma'}=\sum_{i=1}^N d^\dagger_{i,\alpha\sigma}d^\dagger_{i,\beta\sigma'}$. 
While the values of $\chi$ depend on the number of Matsubara frequencies used in the solution of the Bethe-Salpeter equation, the divergent points (zero crossings of $1/\chi$) do not. A smooth crossing of zero implies a possible second order transition to the corresponding ordered state. 
In the spin-rotationally invariant case, the $1\!\!\uparrow$-$2\!\!\uparrow$ and $1\!\!\uparrow$-$2\!\!\downarrow$ pair susceptibilities diverge at the same point, reflecting the degenerate spin-triplet components.

To locate the metal-insulator transitions, we also plot the quantity $-\beta G(\beta/2)$, which at low and fixed temperature should be proportional to the density of states at the Fermi level. It is apparent that the stability region of the superconducting phases end for large $|J|$ at or near these insulator phase boundaries. 
Since the metal-insulator transitions are first order at $T=0.02$ (below the critical end-point),  
the pairing susceptibility does not diverge in their vicinity, i.e., $1/\chi$ jumps from finite negative to finite positive values.
We focus here on the stability region of the metal, so that the phase transition points correspond to $J_{c2}$. If the DMFT calculations were instead started from an insulating solution, a different transition point $J_{c1}$ would be found. We mark $J_{c2}$ by solid black lines in Fig.~\ref{fig_half} and $J_{c1}$ by dashed lines ($J<0$ only). Our calculations indicate that the SC phase exists (down to the lowest accessible temperatures) entirely within the coexistence region of the first-order metal-insulator transition for $U=4$, while it extends beyond the coexistence region for $U=2$.

We also plot the inverse susceptibilities for AFM and AOO, which indicate antiferromagnetic order for $J\gtrsim 0$ and orbital order for $J \lesssim 0$. 
The CO susceptibility diverges near the transition to the double-paired state. From the divergent points of these susceptibilities, we obtain the temperature dependent phase boundaries in Fig.~\ref{fig_half}. 
Note that the analysis of the susceptibilities does not allow us to determine whether and where the transition from one long-range ordered state to another occurs. Therefore, in this half-filled phase diagram, it is not clear if the superconducting phases contained within the antiferromagnetically or antiferro orbital ordered region really emerges. 
Physically, the entropy is lowered in the presence of diagonal order and superconductivity is unlikely to occur.
However, as we will discuss in Sec.~\ref{sec:38}, the superconducting phases extend beyond these ordered regions away from half-filling.

\subsection{Intraorbital spin-singlet superconductivity and paired Mott state}

The SC phase is analogous to the unconventional superconductivity discussed in connection with alkali-doped fullerides,\cite{Capone_2002,nomura_C60_paper,nomura_C60_review} 
in the sense that it involves intraorbital spin-singlet electron pairs stabilized by a negative $J$.
However, a direct comparison is not possible since the fullerides are half-filled three-orbital systems. 
In particular, the paired Mott state located next to the SC phase has a different character in half-filled two- and three-orbital models, because of the odd number of electrons in the latter case. 

The attraction for the SC pairs 
increases as $|J|$ becomes larger.
On theoretical grounds, when $|J|$ is small  we expect that the increasing charge fluctuations (increasing kinetic energy) at small $U$ inhibit the pair formation and destabilize the SC phase, while in the large-$U$ regime, $T_c$ will decrease with increasing $U$, due to a loss of coherence, even though the strength of the effective attraction 
increases\footnote{The maximum $T_c$ for this superconducting state is slightly lower for $U=4$ than for $U=2$, which may be due to the shift of the PM phase boundary to smaller negative values of $J$. 
} 
(note that the renormalization of the electronic kinetic energy by $U$ drives the system into the strong coupling regime\cite{Capone_2002,nomura_C60_paper,nomura_C60_review}).
The physics in this regime would be similar to the large-$|U|$ attractive Hubbard model, where $T_c$ is controlled by the superfluid stiffness.\cite{PhysRevB.72.235118}  

In the rotationally invariant case, the $|J|$ values needed to drive the transition into the spin-singlet superconductor and Mott insulator are about a factor of two smaller than in the Ising case. This can be understood as resulting from the stabilization of the intra-orbital pairs by the pair-hopping term (for $J<0$, the pair-hopping favors the doubly occupied orbital, while the spin-flip term becomes irrelevant). The same effect also explains the enhanced maximum $T_c$ in the rotationally invariant model, compared to the density-density case.\cite{nomura_C60_paper,nomura_C60_review}

As we will discuss in Sec.~\ref{sec:orbitalfreezing}, orbital fluctuations play an important role in inducing the instability to the SC phase. However, in the half-filled case at finite temperature, the transition into the paired Mott phase is strongly first order. 
Therefore, the SC phase is cut off by the first order transition before we see a further development of the orbital fluctuations. 
If we could destabilize the paired Mott phase in some way, the SC region might expand and $T_c$ might increase. 
This point will be revisited in Sec.~\ref{sec:38}.

\subsection{
Interorbital spin-triplet superconductivity and Mott insulator}

\begin{figure*}
	\includegraphics[width=8.7cm]{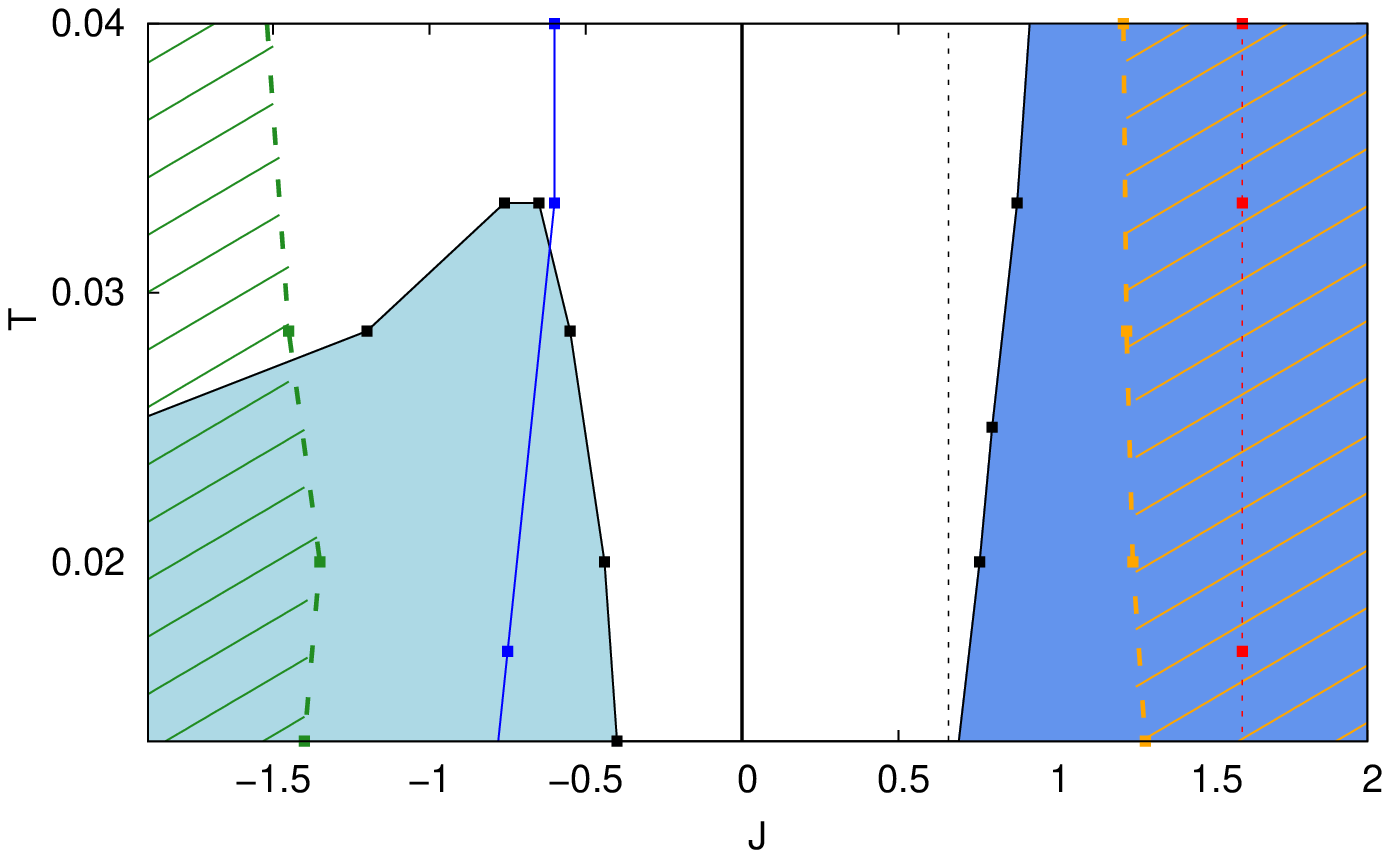}\hfill
	\includegraphics[width=8.7cm]{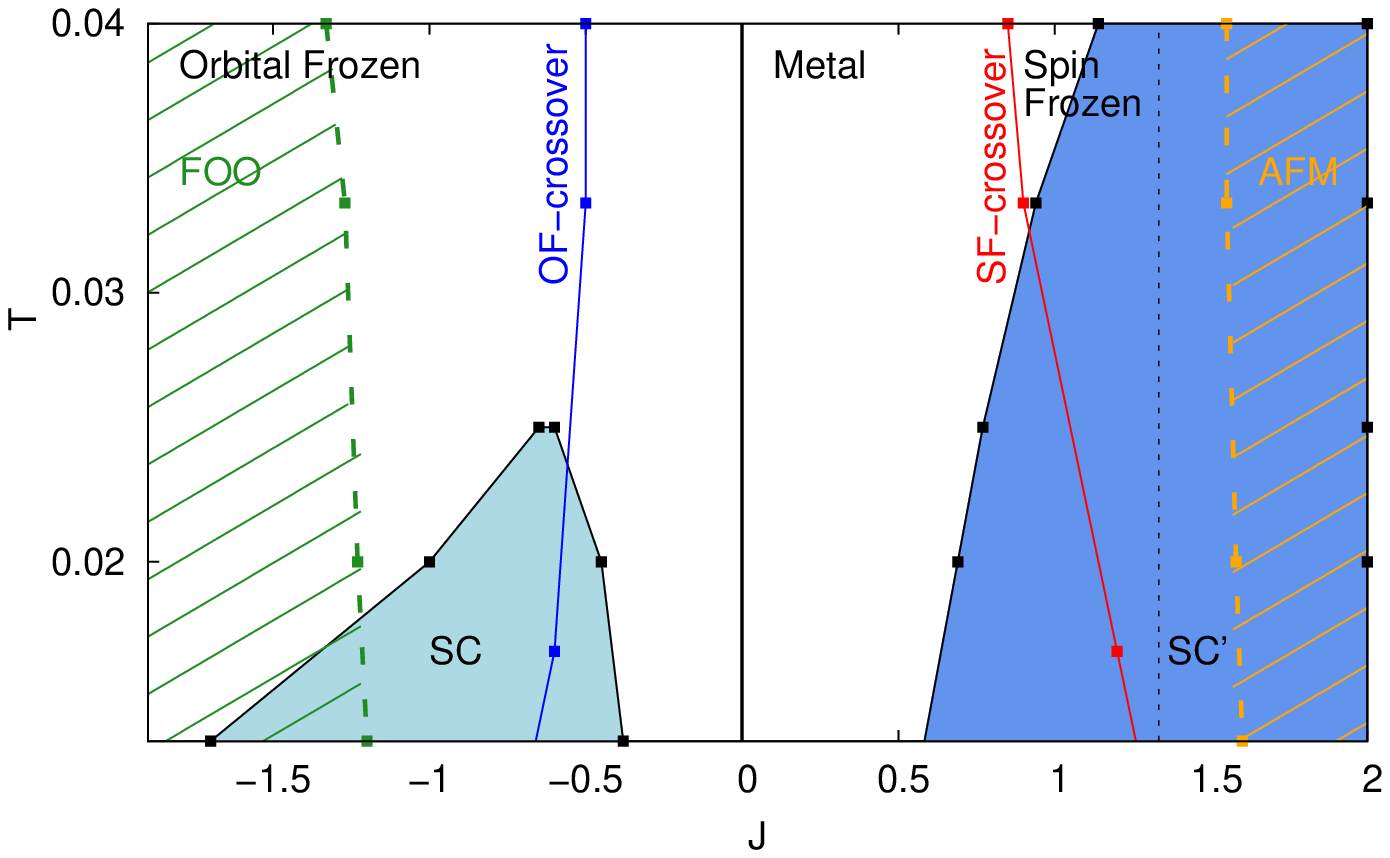}
	\includegraphics[width=8.7cm]{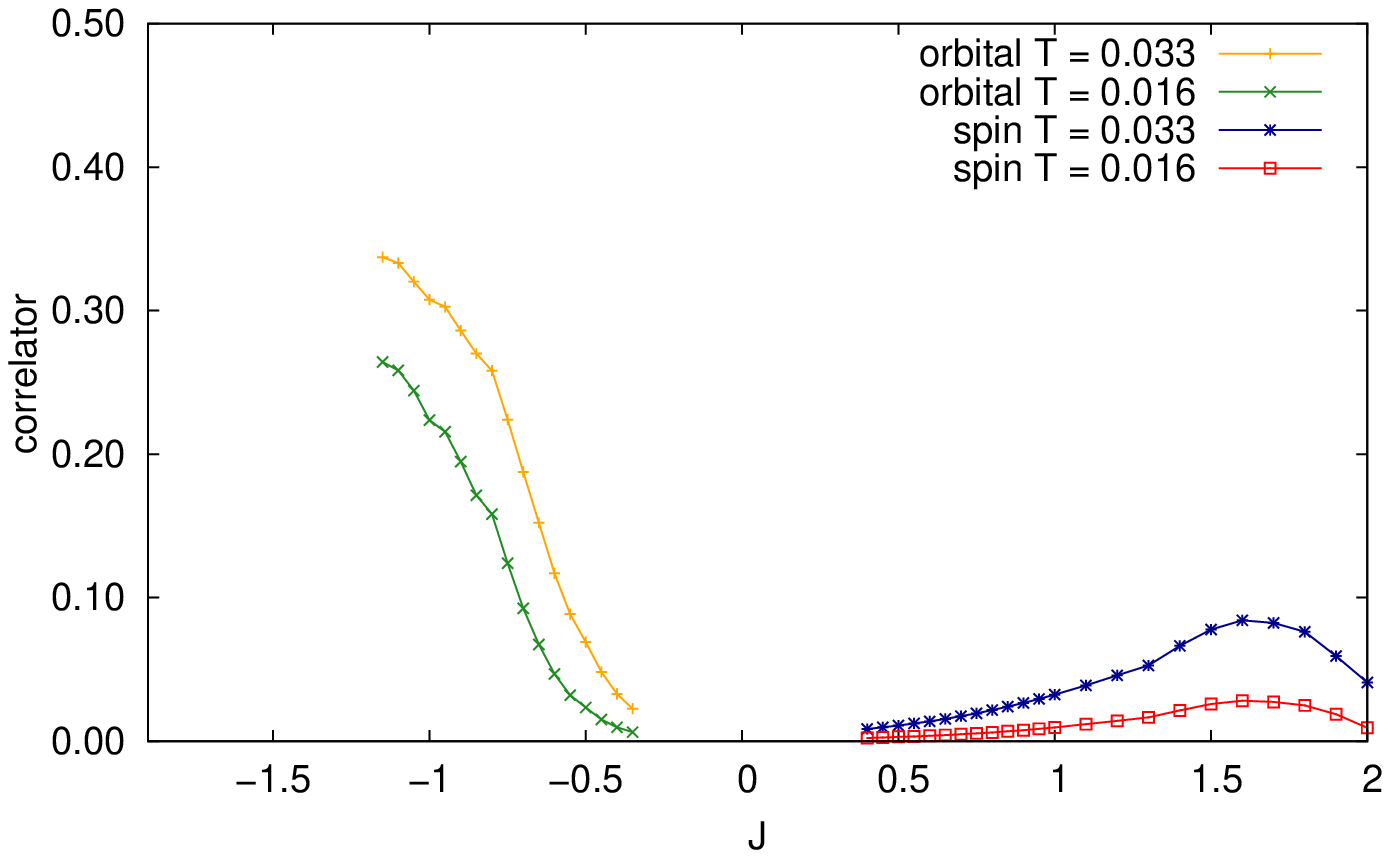}\hfill
	\includegraphics[width=8.7cm]{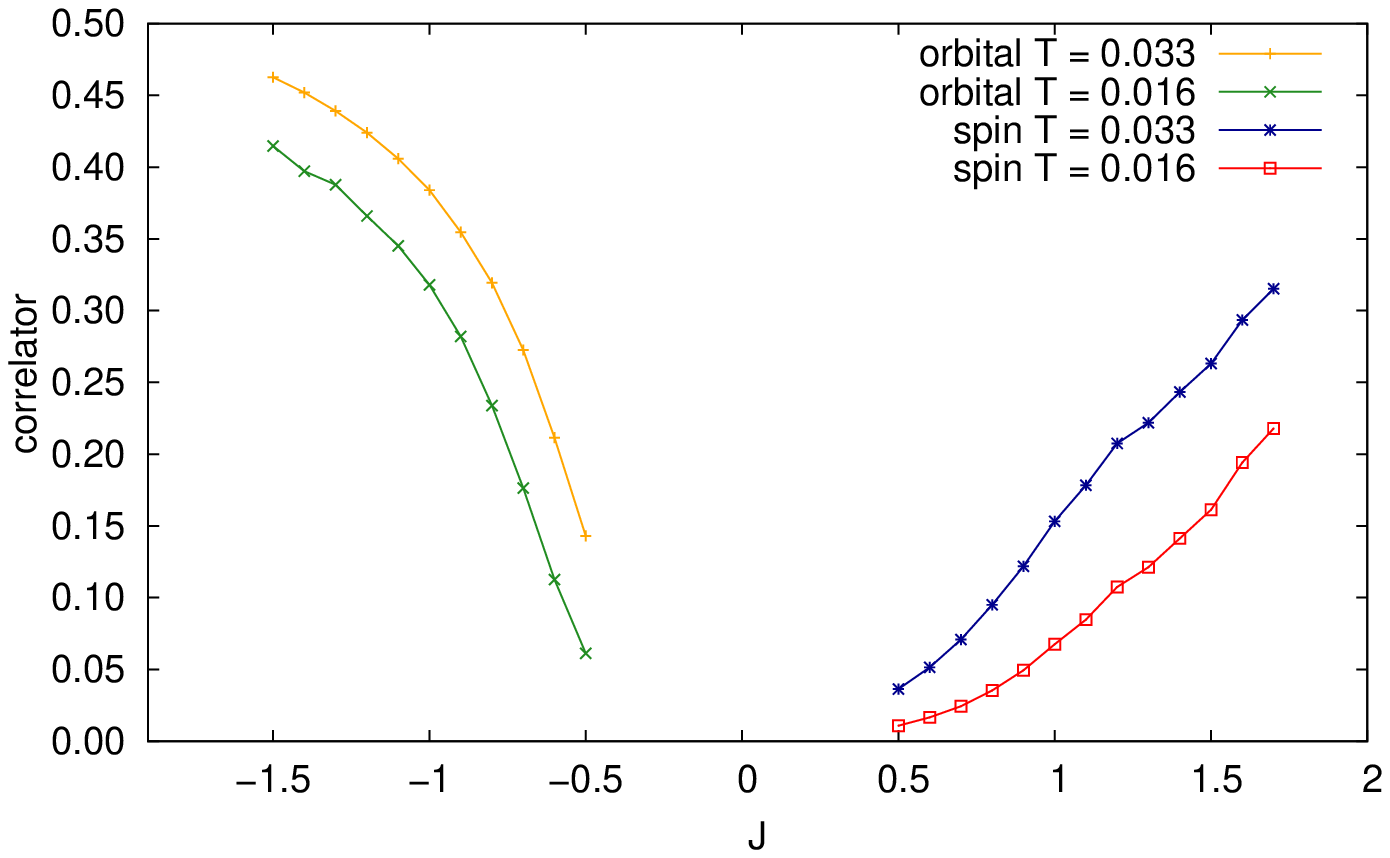}
	\caption{\label{fig_3/8}
	Top panels: Phase diagrams for filling $n_\text{tot}=1.5$ and density-density interactions. 
	The left panel shows the result for $U=2$ and the right panel for $U=4$.
	 Solid and dotted red lines indicate the maxima of the local spin fluctuations, while solid blue lines indicate the maxima of the orbital fluctuations. In the absence of long-range order, the solid lines mark the crossover from a Fermi liquid metal to a spin-frozen ($J>0$) and orbital-frozen ($J<0$) bad metal, respectively. 
	The green hashed region corresponds to ferro orbital order (FOO) and the yellow hashed region to antiferromagnetic order (AFM).
	Bottom panels: Long-time values of the spin-spin and orbital-orbital correlators, $\langle S_z(0)S_z(\beta/2)\rangle$ and $\tfrac14\langle (n_1-n_2)(0)(n_1-n_2)(\beta/2) \rangle$, for indicated temperatures.
}
\end{figure*}

In the $U=2$ case, most of the SC' phase, and also the double-paired state, occur in the 
``unusual" region $J>U/3$ (indicated by the dashed vertical line in the left panels of Fig.~\ref{fig_half}). Note that, when $J>U/3$, the interorbital same-spin interaction $U'-J$ becomes attractive, which explains the high $T_c$. 
At low temperature ($T\lesssim 0.02$), in the density-density approximation, the symmetry breaking occurs within the 
usual 
regime, i.e. for parameters where all the intra- and inter-orbital interaction terms are repulsive. 
 For $U=4$, the entire superconducting phase and the Mott insulator phase boundary are located within the 
 usual 
 $J$ regime ($J<U/3$). 
The orbital singlet, spin triplet nature of the SC' phase is a consequence of the fact that $J >0$ favors high-spin states.

In stark contrast to the case of the spin-singlet superconductivity, the spin-triplet superconducting phase in the $J>0$ region is destabilized by the rotationally invariant interaction. As in the 3-orbital case discussed in Ref.~\onlinecite{Hoshino2015}, we thus find that longitudinal spin fluctuations favor the spin-triplet pairing, while fluctuations among the three degenerate triplet states destroy the coherence. This is consistent with the conclusions reached from the Eliashberg analysis of a single band model.\cite{Monthoux1999}

\section{Results for 3/8 filling}
\label{sec:38}

\subsection{Phase diagrams}

We next compute the diagrams for the 3/8 filled system (1.5 electrons in 2 orbitals) with Ising anisotropy, see Fig.~\ref{fig_3/8}. Away from integer filling, there is no Mott phase, and both the spin singlet and spin triplet superconducting phases expand into the large-$|J|$ region.  We do however find a possible transition into an antiferromagnetic and ferro orbital ordered phase at large positive and negative $J$, respectively. (At larger interactions, e.g. $U=8$, there is an instability to ferromagnetic  order on the $J>0$ side, consistent with the results in Ref.~\onlinecite{Hoshino2016},  
but we do not map out the phase diagram for this parameter regime here.)

\subsection{Spin freezing and its relation to interorbital spin-triplet superconductivity}
\label{sec:spinfreezing}

For $U=2$ and $\beta \le 75$, 
as in the case of half-filling, the spin-triplet superconducting phase is stabilized in the 
unusual 
region $J>U/3$. For $U=4$, we find a SC' region also in the 
usual 
parameter range. In analogy to the 3-orbital case discussed in Ref.~\onlinecite{Hoshino2015}, the spin-triplet superconductivity is  
enhanced 
in the crossover region to the spin-frozen metallic phase at large $J$. 
To illustrate this, we indicate the $J$-values corresponding to maximal local spin-fluctuations 
\begin{equation}
\Delta \chi^\text{spin}_\text{loc}=\int_0^\beta d\tau ( \langle S_z(\tau)S_z(0)\rangle - \langle S_z(\beta/2)S_z(0)\rangle )
\end{equation}
by a red line.  
In the usual parameter regime, where the long-time correlator $\langle S_z(\beta/2)S_z(0)\rangle$ becomes large and essentially temperature-independent in the spin-frozen metal, 
this maximum provides a useful definition of the spin-freezing crossover line. 
Specifically, for $U=4$, the emergence and enhancement \footnote{That the orbital-singlet, spin-triplet superconductivity is enhanced near the spin-freezing crossover line can be deduced from the concave shape of the phase-boundary.} of the SC' phase in the region of maximal local spin fluctuations suggests 
  that these fluctuations are 
responsible for the pairing, an interpretation which has been supported with further numerical data and analytical arguments in Ref.~\onlinecite{Hoshino2015}. We also note that the slope of the spin-freezing crossover line 
in the 
usual
region of the $T$-$J$ phase diagram indicates that the 
spin-frozen 
 region is stabilized by temperature, which is due to the large entropy of the disordered local moments. 

In the unusual
 region $J>U/3$, attractive interactions between same-spin electrons contribute to the pairing. 
While for $U/3 < J \lesssim U$, 
the local spin-fluctuations might still play a role in the pairing, for $J \gtrsim U$, the local 
moment 
is quenched and the direct attractive interaction drives the superconductivity. 
In the $U=2$ case (left panel of Fig.~\ref{fig_3/8}), the maximum local moment fluctuations occur along the dashed red line, in the unusual regime and quite close to $J=U$. Hence, because of the quenching of the moments on the large-$J$ side of this line, there is 
no proper spin-freezing, as evidenced by the large temperature dependence of the long-time correlator $\langle S_z(\beta/2)S_z(0)\rangle$, which is illustrated in the lower panel.

\subsection{Orbital freezing and its relation to intraorbital spin-singlet superconductivity}
\label{sec:orbitalfreezing}

\begin{figure}
	\begin{minipage}[b]{.24\textwidth}
	\includegraphics[width=4.3cm]{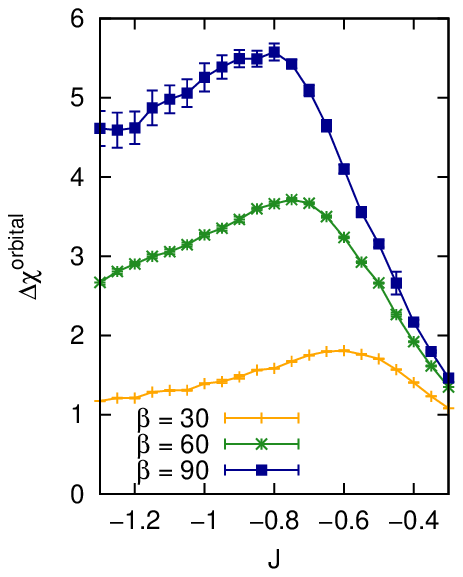}\hfill
	\end{minipage}\hfill
	\begin{minipage}[b]{.24\textwidth}
	\includegraphics[width=4.3cm]{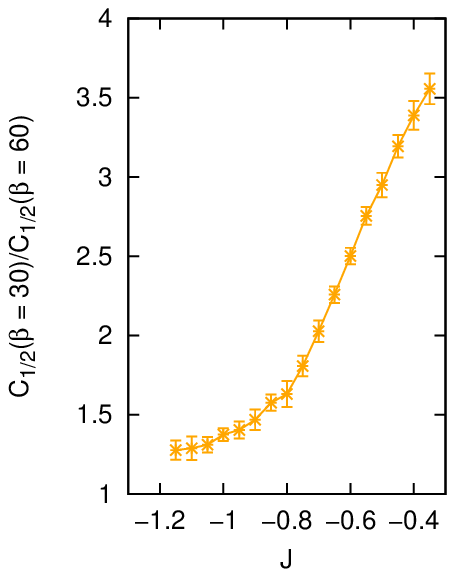}
	\end{minipage}\\
	\caption{\label{fig_ofluct}
	Orbital freezing in the Ising anisotropic model with $U=2$. Left panel: Local orbital fluctuations measured by the correlation function (\ref{chi_orbital}) for indicated values of the inverse temperature. The maxima of these curves define the orbital freezing crossover line displayed in Fig.~\ref{fig_3/8}. Right panel: ratio of two orbital correlation functions measured at inverse temperature $\beta=30$ and $60$, indicating a crossover from Fermi liquid metal to orbitally frozen metal with increasing negative $J$. 
}
\end{figure}

An interesting question is if the SC dome in Fig.~\ref{fig_3/8} is also related to some crossover phenomenon within the $J<0$ metallic phase. To investigate this issue, we plot in the left panel of Fig.~\ref{fig_ofluct} the orbital fluctuations defined by 
\begin{align}
\Delta \chi^\text{orbital}_\text{loc}=& \frac{1}{4} \int_0^\beta d\tau ( \langle (n_1-n_2)(\tau)(n_1-n_2)(0)\rangle\nonumber\\
&\hspace{10mm} - \langle (n_1-n_2)(\beta/2)(n_1-n_2)(0)\rangle ).
\label{chi_orbital}
\end{align}  
(The factor $1/4$ has been added to obtain spin and orbital fluctuations of the same order of magnitude in Fig.~\ref{fig_half}.)
The orbital fluctuations show a temperature dependent maximum in the $J$ region of the SC dome, as illustrated by the 
blue line in Fig.~\ref{fig_3/8}. 
As in the case of the interorbital spin-triplet superconducting state in the $J>0$ region, the intra-orbital spin-singlet superconductor for $J<0$ is induced by strong local fluctuations, which in this case are {\it orbital fluctuations}. The underlying orbital freezing phenomenon is illustrated in the right panel of Fig.~\ref{fig_ofluct}, which shows the ratio of two orbital correlation functions measured at $\tau=\beta/2$, 
\begin{equation}
C_{1/2}(\beta)\equiv \tfrac14 \langle (n_1-n_2)(\beta/2)(n_1-n_2)(0)\rangle,
\end{equation}
for $\beta=1/T$ and $1/(2T)$. In complete analogy to the spin correlation functions analyzed in Ref.~\onlinecite{Werner2009} one observes a crossover from a value of about $1$ (indicating orbital freezing), through $\approx 2$ in the regime of maximal orbital fluctuations (orbital freezing line) to $4$ in the Fermi liquid metal state. 

To show direct evidence for orbital freezing, we also measured the long-time values of the orbital correlation function $\tfrac14 \langle (n_1-n_2)(\beta/2)(n_1-n_2)(0)\rangle$ as a function of $J$. The results are shown in the bottom panels of Fig.~\ref{fig_3/8}. 
The fact that these long-time correlations become finite and weakly dependent on temperature is consistent with the freezing of the orbital moment on the large-$|J|$ side of the crossover line. 
It is interesting to note that the orbital fluctuations also grow in the half-filled system close to the PM insulator. As shown in the narrow panels below the phase diagrams of Fig.~\ref{fig_half}, $\Delta \chi_\text{loc}^\text{orbital}$ exhibits an 
upturn 
in the SC region, but no maximum (as in the 3/8 filled system), because the growth of the orbital fluctuations is cut off by the first order transition into the PM state. This also suggests that the $T_c$ of the SC phase 
could be enhanced if it were possible to destabilize the insulator. Similarly, the spin fluctuations grow in the SC' region as one approaches the half-filled MI in the 
usual 
$J$ regime.

\section{Mapping from $J<0$ to $J>0$}

Here we show that the intraorbital spin-singlet pairing and interorbital spin-triplet pairing can be discussed in a unified manner for the Ising anisotropic case.
Our strategy is to exploit the local transformation defined by
\begin{align}
\begin{pmatrix}
d_{i,1\downarrow} \\
d_{i,2\uparrow}
\end{pmatrix}
\longrightarrow
\begin{pmatrix}
0&1\\
1&0
\end{pmatrix}
\begin{pmatrix}
d_{i,1\downarrow} \\
d_{i,2\uparrow}
\end{pmatrix}.
\label{eq:transf}
\end{align}
While this transformation does not change the kinetic energy term, the density-density interaction is modified as
\begin{align}
H_\text{int}^\text{dens} 
&\longrightarrow 
\sum_\alpha \tilde U n_{\alpha\uparrow} n_{\alpha\downarrow} 
+\sum_{\sigma} \tilde U' n_{1\sigma} n_{2\bar\sigma} 
+\sum_{\sigma} \tilde U'' n_{1\sigma} n_{2\sigma}.
\end{align}
In the original $H_\text{int}^\text{dens} $ the interaction parameters are $U$, $U'=U-2J$ and $U''=U-3J$, while  
for the transformed interaction, the parameters are given by $\tilde U= U-3J$, $\tilde U' = U-2J$ and $\tilde U''=U$.
When $J$ is negative, $\tilde U$ becomes larger than $\tilde U'$. 
Although we have the relation 
\begin{equation}
\tilde U'=\tilde U+J = \tilde{U} - |J|, 
\label{interactions_mapped}
\end{equation}
the qualitative behavior may be expected to be similar to that for $\tilde U'=\tilde U+2J = \tilde{U} - 2|J|$, since the relative order in the magnitude of the local configuration energies remains unchanged.
Thus, the model with negative $J$ is effectively mapped onto the model with positive $J$ by Eq.~\eqref{eq:transf}.

Next, we perform the same transformation on the intraorbital spin-singlet pair amplitude and orbital moment.
The results are
\begin{align}
d^\dagger_{i,1\uparrow} d^\dagger_{i,1\downarrow}
&\longrightarrow d^\dagger_{i,1\uparrow} d^\dagger_{i,2\uparrow},
\\
d^\dagger_{i,2\uparrow} d^\dagger_{i,2\downarrow}
&\longrightarrow d^\dagger_{i,1\downarrow} d^\dagger_{i,2\downarrow},
\\
\sum_{\sigma} (n_{i,1\sigma} - n_{i,2\sigma})
&\longrightarrow 
\sum_{\alpha} (n_{i,\alpha\uparrow} - n_{i,\alpha\downarrow}), 
\end{align}
i.~e., SC is transformed into SC', and AOO into AFM (FOO into FM).
This is the reason why we observe a similarity between the $J>0$ and $J<0$ cases, as shown in Fig.~\ref{fig_half}.
At 3/8 filling, for $U=2$ and $U=4$ (Fig.~\ref{fig_3/8}), where AFM and FOO appears in the $J>0$ and $J<0$ region, respectively, 
the result seems to be inconsistent with the mapping. However, let us note again that for larger interactions ($U=8$), there is FM order on the $J>0$ side of the phasediagram, and FOO on the $J<0$ side, in agreement with the above argument. 
The  
deviation at weaker interactions occurs because the mapping is not exact, i.e., while the order of the interaction strenghts for intra-orbital and inter-orbital same/opposite spin interations is correctly reproduced by the mapping, the ratios between the different couplings are modified (see Eq.~(\ref{interactions_mapped})).

We finally comment on the spin isotropic case.
The spin-flip and  pair-hopping terms are transformed as
\begin{align}
H_\text{int}^\text{sf-ph}  \longrightarrow
J (d_{2\uparrow}^{\dagger} d_{1\downarrow}^{\dagger} d_{2\downarrow} d_{1\uparrow} + d_{1\downarrow}^{\dagger} d_{2\downarrow}^{\dagger} d_{1\uparrow} d_{2\uparrow}) + {\rm h.c.}
\label{pair_flip}
\end{align}
Here, the second term on the right-hand side is a new term which is not included in the original Hamiltonian.
Hence in this case the qualitative similarity between the models with $J>0$ and $J<0$ is not guaranteed.
Indeed, the effect of $H_\text{int}^\text{sf-ph}$ on the spin-singlet and spin-triplet superconducting phase is different (Fig.~\ref{fig_half}): It increases the $T_c$ for SC but decreases $T_c$ for SC'. 
The pair hopping term stabilizes the intra-orbital pairs for $J<0$, and hence the second term in Eq.~(\ref{pair_flip}) enhances spin-triplet pairs for $J>0$. However, such a ``double-spin-flip" term is not included in the original Hamiltonian and we instead have the ordinary spin-flip term, which leads to a decrease of $T_c$.\cite{Monthoux1999}

\section{Conclusions and outlook}
\label{sec:outlook}

We have presented a systematic study of the ordered phases and crossovers in two-orbital Hubbard models. At half-filling, for Hund coupling $J>0$, we found an orbital-singlet spin-triplet superconducting phase next to a high-spin Mott insulator, in agreement with previous work on three orbital models.\cite{Hoshino2015} For small $U$, this phase occurs in the 
``unusual" 
region $J>U/3$, with attractive same-spin intra-orbital interactions, and is stabilized at low temperature up to $J\approx U$, where the transition into a double paired low-spin state takes place. The instability to antiferromagnetic order, however, occurs at substantially higher temperatures. At 3/8 filling, where the insulating phases disappear and antiferromagnetic order is suppressed, we find an extended 
spin-triplet superconducting region 
outside the antiferromagnetic phase. 
For 
usual
Hund couplings ($0<J<U/3$) this superconducting phase is intricately connected to the spin-freezing crossover which occurs in the disordered metal phase, and which is marked by large fluctuations of the local moments. These fluctuating moments have been shown in Ref.~\onlinecite{Hoshino2015} to cause the spin-triplet pairing. The spin-triplet superconductivity in the 
unusual
region $J>U/3$   
is enhanced by direct attractive interactions between same-spin electrons in different orbitals. 

The main new result of this study is that similar physics also appears in the model with $J<0$, albeit with the role of orbital and spin degrees of freedom interchanged. For sufficiently attractive $J$, an instability to an intra-orbital spin-singlet superconducting phase appears next to a paired Mott insulator. At half-filling, this superconducing phase is contained within an antiferro orbitally ordred region, but at 3/8 filling, where the paired Mott insulator disappears and orbital order is suppressed, there is an extended spin-singlet superconducting region 
outside the orbitally ordered phase.
The peak of the superconducting dome coincides with a maximum in local orbital fluctuations. These strong orbital fluctuations mark an orbital freezing crossover which occurs within the disordered metal phase and separates a Fermi liquid region with fast decaying orbital correlations from an orbital-frozen non-Fermi liquid region. 

We used symmetry arguments to connect the spin-triplet and spin-singlet superconducting phases and the underlying spin and orbital freezing phenomena, thereby establishing a tight connection between the two types of unconventional superconductors, which are relevant, respectively, for the theoretical understanding of strontium ruthenates ($J>0$) and alkali-doped fullerides ($J<0$). Concerning the fullerides, we have made the relevant observation that the orbital fluctuations also exhibit an increase in the spin-singlet superconducting region of the half-filled system. The growth of these fluctuations is however cut off by the transition into the paired Mott insulator. The $T_c$ of the half-filled model could thus be enhanced by increasing the orbital fluctuations, increasing the effective attractive $J$, and at the same time preventing a transition into the paired Mott state, an interesting result in connection with the recently observed light-enhanced superconductivity in K$_3$C$_{60}$.\cite{Mitrano2015}

\acknowledgements
The simulations were run on the BEO04 cluster at the University of Fribourg, using a code based on ALPS.\cite{Albuquerque2006} KS was supported by SNSF grant No.~200021-140648,  YN by the Consolidator Grant CORRELMAT of the European Research Council (project number 617196), and PW by ERC starting grant No. 278023.

\bibliography{hw}

\end{document}